\documentclass[conference,a4paper]{IEEEtran}
\IEEEoverridecommandlockouts
\usepackage[english]{babel}
\usepackage[utf8x]{inputenc}
\usepackage{cite}
\usepackage{amsmath,amssymb,amsfonts}
\usepackage{amsthm}
\usepackage{algorithmic}
\usepackage{algorithm}
\usepackage{graphicx}
\usepackage{textcomp}
\usepackage{xcolor}
\usepackage{stfloats}
\usepackage{multirow}
\usepackage{verbatim}
\usepackage{textcomp}
\def\BibTeX{{\rm B\kern-.05em{\sc i\kern-.025em b}\kern-.08em
    T\kern-.1667em\lower.7ex\hbox{E}\kern-.125emX}}

\newtheorem{lemma}{Lemma}

\begin{document}
\allowdisplaybreaks

\title{
%Network Orchestration in Mobile Networks using a Synergy between Model-driven and AI-based Techniques
Network Orchestration in Mobile Networks via a Synergy of Model-driven and AI-based Techniques
}

\author{\IEEEauthorblockN{Yantong Wang}
	\IEEEauthorblockA{
		Department of Engineering\\ King's College London\\
		London, WC2R 2LS, U.K. \\
		yantong.wang@kcl.ac.uk}
	\and
	\IEEEauthorblockN{Vasilis Friderikos}
	\IEEEauthorblockA{
		Department of Engineering\\ King's College London\\
		London, WC2R 2LS, U.K. \\
		vasilis.friderikos@kcl.ac.uk}
	\thanks{This is an accepted version of ComNet2020 \textcopyright2020 IEEE. Personal use of this material is permitted. Permission from IEEE must be obtained for all other uses, in any current or future media, including reprinting/republishing this material for advertising or promotional purposes, creating new collective works, for resale or redistribution to servers or lists, or reuse of any copyrighted component of this work in other works.}
}
\maketitle

\begin{abstract}
As data traffic volume continues to increase, caching of popular content at strategic network locations closer to the end user can enhance not only user experience but ease the utilization of highly congested links in the network. A key challenge in the area of proactive caching is finding the optimal locations to host the popular content items under various optimization criteria. These problems are combinatorial in nature and therefore finding optimal and/or near optimal decisions is computationally expensive. In this paper a framework is proposed to reduce the computational complexity of the underlying integer mathematical program by first predicting decision variables related to optimal locations using a deep convolutional neural network (CNN). The CNN is trained in an offline manner with optimal solutions and is then used to feed a much smaller optimization problems which is amenable for real-time decision making. Numerical investigations reveal that the proposed approach can provide in an online manner high quality decision making; a feature which is crucially important for real-world implementations. 
	%The CNN is trained first with optimal solutions and numerical investigations reveal that the performance can increase by more than 400\% compared to powerful randomized greedy algorithms. To this end, the proposed technique seems as a promising way forward to the holy grail  aspect in resource orchestration which is providing high quality decision making in real time.
	% save 100x running time with less than 5 precent performance loss
\end{abstract}


\section{Introduction}
\label{sec:intro}
%The network has gone through a digitization revolution during last decades. From the Global System for Mobile Communications (GSM) in the second-generation (2G) introducing circuit switched service to the Long Term Evolution (LTE) standard for the forth-generation (4G) generating a full IP packet switched network for all services, not only end users but also terminal devices are connected together, which provides the fundament of Internet of Things (IoT). Based on the key capabilities of 4G, the fifth-generation (5G) proposes higher requirements such as 10x user experienced data rate and 100x area traffic capacity in the enhanced mobile broadband (eMBB) scenario or 0.1x latency in the ultra-reliable and low latency communication (URLLC) scenario compared with current network\cite{series2015imt}. Moreover, the ever-continuing explosive growth of data traffic will be more than 136 exabytes per month in 2024\cite{cerwall2018ericsson}. Both the variety of very diverse requirements in different scenarios and data booming impose great pressure on today's network.
\IEEEPARstart{S}{ince} 2016 when AlphaGo managed to win the 18-times world champion Lee Sedol at the Go match\footnote{www.deepmind.com/alphago-korea}, AI techniques started to attract significant attention from both academia and industry within the ICT sector for a plethora of different diverse complex network management applications both at the control and data plane. Even though we are still in a rather embryonic stage in adopting these types of technologies, the potential benefits of deep learning techniques are significant to tame the underlying complexity of the emerging highly heterogeneous 5G and beyond mobile networks. 
In current work, our attention is on the issue of proactive caching of popular content using an amalgamation of model based techniques and deep learning in the form of convolutional neural networks (CNNs). Caching of popular content is a key technique to enhance  user experience and overall network performance; it has attracted an overwhelming research attention recently \cite{SurveyCaching}. 
%When mobile device changes its access point the supported caching policies can be classified as follows \cite{vasilakos2012proactive}: reactive approach \cite{sourlas2010mobility}, durable subscription\cite{farooq2004performance} and proactive caching\cite{gaddah2010extending}. 
%For reactive approach, the previous access point keeps caching the requested content after mobile user departure and restarts transmission to the new proxy when re-connection. In durable subscription, not only the previous connected point but all potential access points should keep caching the content in case of information loss. Obviously, the reactive approach suffers from delay for the re-transmission after handover and the durable subscription pay significant cost for memory usage. Nevertheless, the proactive caching makes a trade-off between available caching storage and latency by hosting the information in a subset of all candidates.
%%Compared with other caching policies, proactive caching makes a trade-off between storage and transmission. 
%%Generally, there are two inter-linked  sub-problems in proactive caching that need to be tackled; firstly, the issue of where to cache and secondly what to cache in each selected location. In this paper, we focus mainly on the first problem which relate on choosing the edge clouds for  hosting popular content. For the purpose of deriving optimal decision making for content placement a popular approach is based on using mixed integer linear programming (MILP) formulations. 
Proactive caching has been mainly considered using (mixed) integer linear programming (MILP) formulations that take into account resource availability in different edge clouds and communication cost \cite{wang2019proactive}. However, we end up with  $NP$-hard\cite{zappone2019wireless} optimization problems which cannot be deemed suitable to support real-time decision making.
%Though 4G works well for today's network, it still has limitations when considering future use cases and business model. With the advent of the fifth-generation (5G), the "one-sizes-fits-all" paradigm in current network is broken and a variety of very diverse requirements in different scenarios are taken into consideration\cite{series2015imt}. 

%Recently, deep learning (DL) technology, a member of the broader family of machine learning, attracted significant attention from both academia and industry. DL is based on artificial neural networks but has multi hidden layers in between the input and output layer and the applications are not only limited to speech or image recognition, but also expended to a diverse set of domains such as self-driving cars, medical diagnosis and playing games such as Go\cite{sze2017efficient}. 
Deep learning (DL) frameworks are gaining significant attention in network research and already DL techniques have been used for resource management, routing and other network related operations \cite{sun2019application}, \cite{sze2017efficient}, \cite{cheng2017mobile}. 
%Undeniably, these AI-based technologies, like DL, will play an important role in future network design and operation. The reason comes from two main aspects: i) the explosive growth of traffic data in future network provides large datasets for DL training\cite{cheng2017mobile}; ii) the improvement of hardware allows for practical implementation AI-based approaches, for example, the graphics processing units (GPUs) execute DL at orders of magnitude faster than traditional Central Processing Units (CPUs). Compared to some other deep learning architectures, Convolutional Neural Network (CNN) has the feature of weight-sharing, which means the same set of weights are used in the processing. Additionally, recent results show that a CNN can exceed human-level accuracy in image recognition \cite{sze2017efficient}.
%All these positive factors inspire us: could we copy the success of CNN in image classification to the area of caching placement?
DL techniques has also made their way into the caching problems \cite{chen2019artificial}. However, most of them focus on the content popularity prediction \cite{tanzil2017adaptive}, \cite{lei2019learning} and there are few works dealing with the problem of finding optimal caching locations. Closely related works can be find in \cite{lee2018deep} and \cite{lei2017deep}, where the former trains a deep neural network (DNN) for solving linear programming problem and the latter apply DNN to answer linear sum assignment problem. As aforementioned, the caching allocation problem belongs to $NP$-hard family and it is harder than solving linear programming or linear sum assignment problem. 
In this paper, the principal idea is representing the proposed optimization problem as a grayscale image which is then used to train a deep CNN with optimal solutions.
Even though caching problems belong to the general class of $NP$-hard problems, this relate to the worst case complexity. 
In other words, the difficulty of solving these type of problems are considered under their hardest case and in reality many small to medium search space instances can be solved very efficiently \cite{nowak2018revised}. 
% In other words, these type of problems are often described to be as hard as their hardest instance. Many hard problems can be efficiently solved for a large subset of inputs. 
It is this feature and characteristic of the problem that we aim to explore with the proposed CNN.
In that respect the CNN provides an estimation of suitable edge cloud locations for hosting popular content that can be used for real time decision making. As will be discussed in the sequel, the proposed technique manages to provide better decision making than nominal real time techniques that are based on greedy heuristics. Hence the proposed framework can be considered as an amalgamation between model based and data-driven techniques. 

\section{System Model}
\label{sec:model}
%\textit{prove NP-hard?}
% In this section we generate the MILP model for proactive caching placement problem, which is the step from network to math model in both Figure \ref{fig:training} and \ref{fig:testing}.

A mobile network is represented with a graph (undirected) $\mathcal{G}=\{\mathcal{V},\mathcal{L}\}$, in which $\mathcal{V}$ represents the nodes in the network and $\mathcal{L}$ denotes the communication links. A subset  $\mathcal{E} \subseteq \mathcal{V}$ of the network nodes are the edge clouds (ECs), or content routers (as commonly named).
%\footnote{The term \textit{edge cloud} and \textit{content router} are used interchangeably in the rest of the paper}. 
By $\mathcal{A} \subseteq \mathcal{V}$, the set of access routers (ARs) is that mobile users connect to as they change point of attachment due to their mobility. Hereafter we assume that mobility information is readily available using historical data that an operator can explore. Also note that the  following might hold $\mathcal{E} \cap \mathcal{A} \neq \emptyset$, which means a subset of access routers can be deemed suitable to cache popular content.

In the sequel a salient assumption is that there is one flow request per user and $k\in\mathcal{K}$ is defined as the set of flows routed through the network. Each request $k$ characterized by the following 3-tuple properties: $s_k$, which represents the requested memory space for caching content of flow $k$; $b_k$, the requested data rate of flow $k$ and $p_{ka}$ which denotes the likelihood for flow $k$ to move to AR $a$, where $a\in \mathcal{A}$. With the same token, for each candidate EC $e \in\mathcal{E}$ we denote with $w_e$ the available memory for storage in EC $e$ and for each link $l\in\mathcal{L}$ we denote the remaining capacity as $c_l$. 
%  Vertices are combined with gateway node, access routers and edge clouds/content routers $\mathcal{E}$ and access routers $\mathcal{A}$ and $\mathcal{E}\cap\mathcal{A}=\emptyset$ for simpler analysis.
% 
% Each mobile user/flow $k$ has two requirements: $s_k$ for interested content storage space and $b_k$ for bandwidth.
% 
% Each edge cloud $e\in\mathcal{E}$ is connected with two properties: total storage space $s_e^T$ and remaining space $s_e^R$. So the initial edge cloud utilization can be estimated as $$u_e^I=1-\frac{s_e^R}{s_e^T}$$ and final edge cloud utilization is $$u_e^F=u_e^I+\frac{\sum_{k\in\mathcal{K}}s_k\cdot x_{ke}}{s_e^T}$$
% 
% Similarly, each link $l\in\mathcal{L}$ has properties: total link bandwidth $b_l^T$ and remaining bandwidth $b_l^R$. The initial link utilization is measured by $$u_l^I=1-\frac{b_l^R}{b_l^T}$$ and final link utilization after assignment is $$u_l^F=u_l^I+\frac{\sum_{k\in\mathcal{K}}b_k\cdot y_{kl}}{b_l^T}$$
Additionally, $N_{ae}$ is the hop counter between connected point $a$ and caching host $e$ via the shortest path. Expressly, $N_{ae}=0$ iff $a=e$; by $N^T$ the number of hops is represented from access point to data center, whose average value is between 10 and 15 in the current network \cite{van2014performance}; $\alpha$ and $\beta$ are the weight of hosting and transmission cost respectively.
The following 0/1 decision variables are introduced that will lead to the proposed integer programming formulation in the sequel,

\hangafter 1
\hangindent 1em
\noindent
\\$  x_{ke}=
\begin{cases}
1,  &\text{if content for flow $k$ is hosted at EC $e$}  \\
0,  &\text{otherwise}
\end{cases}$
\\
\\$ y_{kl}=
\begin{cases}
1, &\text{if flow $k$ use link $l$} \\
0, &\text{otherwise}
\end{cases}$
\\
\\$ z_{kae}=
\begin{cases}
1, &\text{if flow $k$ connect with $a$ and retrieve the}\\ &\text{cached content from EC $e$}\\
0, &\text{otherwise}
\end{cases}$
\\

The total cost ($TC$) in this paper consist of two pieces: 
\begin{equation}
TC=\alpha\cdot C^C+\beta\cdot C^T
\end{equation}
Specifically, $C^C$ is the hosting cost as follows \cite{vasilakos2012proactive}:
% \footnote{In general case, caching cost contains a denominator related with EC utilization. In this paper we simply the cost as the summary of $x$}
\begin{equation}
C^C=\sum_{k\in\mathcal{K}}\sum_{e\in\mathcal{E}}\frac{x_{ke}}{1-U_e}
\end{equation}
and $U_e$ is the space utilization level at EC $e$. So those ECs with less memory usage take the priority over the others in host selection. The value of $U_e$ relies on how many contents have been allocated to this EC:
\begin{equation}
U_e=\frac{\sum_{k\in\mathcal{K}}s_k\!\cdot\! x_{ke}}{w_e}=\sum_{k\in\mathcal{K}}q_{ke}\!\cdot\!x_{ke},\forall e\in\mathcal{E}
\end{equation}
where $q_{ke}=s_k/w_e, \forall k\!\in\!\mathcal{K}, e\!\in\!\mathcal{E}$, which indicates the effect of caching selection on space utilization level. Then $C^C$ can be expressed as
\begin{equation}
\label{fml:cc}
C^C=\sum_{k\in\mathcal{K}}\sum_{e\in\mathcal{E}}\frac{x_{ke}}{1-\sum_{k\in\mathcal{K}}q_{ke}x_{ke}}
\end{equation}
In that case $C^C$ is the ratio of two linear terms, where the decision variable $x_{ke}$ in the denominator makes it non-linear. So we can apply some integer linear programming solvers if $C^C$ is transformed to a linear function. It is noting that the value of the denominator is positive due to the fact that the memory used for caching should be less than the entire available space, which is a constraint in the sequel. According to \cite{wang2019proactive}, the main trick is introducing auxiliary decision variables $\chi_{ke}=t_e\!\cdot\!x_{ke}$ and we define
\begin{equation}
t_e=\frac{1}{1-\sum_{k\in\mathcal{K}}q_{ke}\!\cdot\!x_{ke}},\forall e\in\mathcal{E}
\end{equation}
This definition is equivalent to the following constraints:
\begin{subequations}
\begin{align}
\label{fml:t1}
&t_e-\sum_{k\in\mathcal{K}}q_{ke}\!\cdot\!x_{ke}t_e=1,\forall e\in\mathcal{E} \\
\label{fml:t2}
&t_e>0, \forall e\in\mathcal{E}
\end{align}	
\end{subequations}
Considering $t_e$ is a fractional variable while $x_{ke}$ is binary, the value of $\chi_{ke}$ is bounded by constraints:
\begin{subequations}
\begin{align}
& \chi_{ke}\!\leq\!t_e,\forall k\!\in\!\mathcal{K},e\!\in\!\mathcal{E}\\
& \chi_{ke}\!\leq\!M\!\cdot\!x_{ke},\forall k\!\in\!\mathcal{K},e\!\in\!\mathcal{E}\\
& \chi_{ke}\!\geq\!M\!\cdot\!(x_{ke}-1)\!+\!t_e,\forall k\!\in\!\mathcal{K},e\!\in\!\mathcal{E}
\end{align}
\end{subequations}
and $M$ is a sufficiently large number. Then we rewrite $C^C$ in the terms of $\chi_{ke}$
\begin{equation}
C^C=\sum_{k\in\mathcal{K}}\sum_{e\in\mathcal{E}}\chi_{ke}
\end{equation}
% The cost of caching is estimated by many previous works as
% \begin{equation*}
%     C^C=\alpha\cdot\sum_k\sum_e\frac{x_{ke}}{1-u_e^F}
% \end{equation*}
% Here in order to reduce the complexity of model and calculation, we ignore the denominator. Then it becomes
% \begin{equation}
%     C^C=\alpha\cdot\sum_k\sum_e x_{ke}
% \end{equation}

$C^T$ denotes the communication cost which is considered as hop count distance from mobile user to the caching host or data center,
\begin{equation}
\begin{aligned}
\label{eq:trans}
&C^T=C^H+C^M=\\
&\sum_{k\in\mathcal{K}}\sum_{a\in\mathcal{A}}\sum_{e\in\mathcal{E}} p_{ka}N_{ae}z_{kae}\!+\!\sum_{k\in\mathcal{K}}(1\!-\!\sum_{a\in\mathcal{A}}\sum_{e\in\mathcal{E}}p_{ka}z_{kae})N^T
\end{aligned}
\end{equation}
where $C^H$ and $C^M$ describes the cost of cache hitting and missing respectively. The connected AR $a$ of user $k$ is not determined but narrated by the moving probability $p_{ka}$ so $C^T$ is the expected value of transmission hops in \eqref{eq:trans}. If the requested content is caching in the connected access point, the transmission cost will not be paid since $N_{ae}=0$; another case is that the mobile user attach in AR and retrieve to EC, then $C^H$ is going to be considered for the transport; otherwise the system will suffer from large transmission cost $C^M$ due to cache missing. 
% Where $C^H$ represents the cache hitting cost when the mobile user (i.e. flow $k$) connects to the AR $a$ and retrieve to EC $e$. $N_{ae}$ is the number of hops on the shortest path between connected AR $a$ and cache hosting EC $e$. Notably $N_{ae}=0$ iff $a=e$. $C^M$ is the cost for 
% flow $k$ 
% missing cache, which means we cannot retrieve the request from flow $k$ to a caching hoster $e$. Then the request will be redirected to the data center. Here we use $N^T$ to represent the number of hops from AR $a$ to central data server. In the current network, the request traverses an average of $10$ to $15$ hops between source and destination \cite{van2014performance}. 
% So the value of $N^T$ is between $10$ and $15$. 
% explain hit and miss cost

Therefore, the proactive caching problem with the aim of providing interested contents allocations to minimize total cost is formulated as follows,
% The objective of this paper is to determine the optimal caching strategy that minimizes total cost which can be formulated as:
\begin{subequations} \label{LP:main}
\vspace{-0.2cm}
	\begin{align}
	\label{LP:obj}
	&\mathop{\min}_{\substack{x_{ke},y_{kl},z_{kae}\\t_e,\chi_{ke}}}\; \alpha\sum_{k\in\mathcal{K}}\sum_{e\in\mathcal{E}}\chi_{ke}\!+\!\beta\Big(\sum_{k\in\mathcal{K}}\sum_{a\in\mathcal{A}}\sum_{e\in\mathcal{E}}\nonumber\\ 
	&\quad p_{ka}N_{ae}z_{kae}\!+\!\sum_{k\in\mathcal{K}}(1\!-\!\sum_{a\in\mathcal{A}}\sum_{e\in\mathcal{E}}p_{ka}z_{kae})N^T\Big)\\
	\textrm{s.t.}\quad
	\label{LP:con1}
	& \sum_{e\in\mathcal{E}} x_{ke}\!\leq\!1, \forall k\!\in\!\mathcal{K} \\
	\label{LP:con2}
	& \sum_{k\in\mathcal{K}} s_k\!\cdot\!x_{ke}\!\leq\!w_e, \forall e\!\in\!\mathcal{E}\\
	\label{LP:con3}
	& \sum_{e\in\mathcal{E}} z_{kae}\!\leq\!1, \forall k\!\in\!\mathcal{K},a\!\in\!\mathcal{A} \\
	\label{LP:con4}
	& z_{kae}\!\leq\!x_{ke}, \forall k\!\in\!\mathcal{K},a\!\in\!\mathcal{A},e\!\in\!\mathcal{E}\\
	\label{LP:con5}
	& \sum_{k\in\mathcal{K}} b_k\!\cdot\!y_{kl}\!\leq\!c_l,\forall l\in\mathcal{L} \\
	\label{LP:con6}
	& y_{kl}\!\leq\!\sum_{a\in\mathcal{A}}\sum_{e\in\mathcal{E}} B_{lae}\!\cdot\!z_{kae}, \forall k\!\in\!\mathcal{K},l\!\in\!\mathcal{L} \\
	\label{LP:con7}
	& M\!\cdot\!y_{kl}\!\geq\!\sum_{a\in\mathcal{A}}\sum_{e\in\mathcal{E}} B_{lae}\!\cdot\!z_{kae}, \forall k\!\in\!\mathcal{K},l\!\in\!\mathcal{L}\\
	\label{LP:con8}
	& t_e\!-\!\sum_{k\in\mathcal{K}}q_{ke}\!\chi_{ke}\!=\!1,\forall e\!\in\!\mathcal{E}\\
	\label{LP:con9}
	& \chi_{ke}\!\leq\!t_e,\forall k\!\in\!\mathcal{K},e\!\in\!\mathcal{E}\\
	\label{LP:con10}
	& \chi_{ke}\!\leq\!M\!\cdot\!x_{ke},\forall k\!\in\!\mathcal{K},e\!\in\!\mathcal{E}\\
	\label{LP:con11}
	& \chi_{ke}\!\geq\!M\!\cdot\!(x_{ke}-1)\!+\!t_e,\forall k\!\in\!\mathcal{K},e\!\in\!\mathcal{E}\\
	& x_{ke},y_{kl},z_{kae}\!\in\!\{0,1\},\forall k\!\in\!\mathcal{K},l\!\in\!\mathcal{L},a\!\in\!\mathcal{A},e\!\in\!\mathcal{E}\\
	& t_e,\chi_{ke}>0,\forall k\!\in\!\mathcal{K}, e\!\in\!\mathcal{E}
	\end{align}
\end{subequations}
Constraint \eqref{LP:con1} forces at most 1 EC to provide caching service for a mobile user. \eqref{LP:con2} considers the storage limitation for each EC and \eqref{LP:con5} represents the bandwidth capacity for individual link. The retrieved path is expressed in \eqref{LP:con3} and \eqref{LP:con4}, where the former enforce imposes the route is unique when flow and AR are determined and the latter guarantee the destination should store the required information. Furthermore, constraint \eqref{LP:con7} indicates that once the path is selected for retrieving, all the links on this route should be taken into consideration and \eqref{LP:con6} describe the contrary case, where $B_{lae}$ is a binary indicator which is set to 1 if the link $l$ locates in the shortest path between AR $a$ and EC $e$ and is set to 0 otherwise. This three dimension matrix can be constructed from network topology.
% shows the relationship between link and path, which could be generated from network topology by defining $$B_{lae}=
% \begin{cases}
% 1, &\text{if link $l$ in shortest path between $a$ and $e$} \\
% 0, &\text{otherwise}
% \end{cases}$$
The constraint \eqref{LP:con8} as well as \eqref{LP:con9}$\sim$\eqref{LP:con11} 
constitute the definition of auxiliary variable $t_e$ and $\chi_{ke}$ respectively, which are used to linearize the objective function \eqref{LP:obj}.

\section{\textcolor{black}{Proposed AI-based Approach}}
\label{sec:DNN}
\subsection{Training Process}
\label{sec:subtraining}
\begin{figure}[htb]
	\centering
	\includegraphics[width=0.48\textwidth]{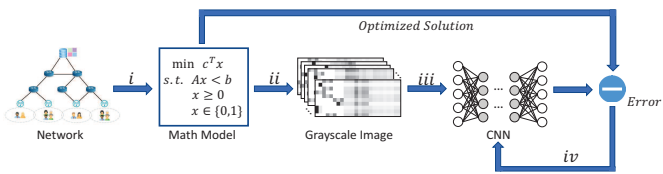}
	\caption{Training Process}
	\label{fig:training}
\end{figure}

The training process mainly comprise four steps as illustrated  in Figure \ref{fig:training}. In step $i$, we construct the mathematical model from the network as represented in Section \ref{sec:model}.

In step $ii$, we generate the grayscale image by extracting the main characters of optimization model \eqref{LP:main}. Depending on the analysis of optimal solutions, we find that once the network topology is fixed, these parameters which relies on the network topology, such as $N_{ae}$ and $B_{lae}$, have very limited contribution to the final allocations. On the contrast, those variables describing the user behavior, resource requirements and current network resource distribution play an important role in the cache assignment. For instance, those ECs, which have more available storage space and more closed to mobile users, have higher probability to be picked as caching host than the others. In our problem, we extract the user moving probability $p_{ka}$, interested content storage requirement $s_k$, required bandwidth $b_k$, available space in each EC $w_e$ and available link capacity $c_l$ as the key parameters of problem \eqref{LP:main} to construct the grayscale image. However, these five variables have different dimensions. For the aim of dimension matching, we use $q_{ke}$ which is introduced in Section \ref{sec:model} as a combination of $s_k$ and $w_e$. Similarly, $r_{kl}$ is defined as the ratio of $b_k$ and $c_l$, indicating the impact of caching selection on link bandwidth utilization level. Consequently, the two-dimensional grayscale image is composed of $p_{ka}$, $q_{ke}$ and $r_{kl}$ whose size is $|\mathcal{K}|\times(|\mathcal{A}|+|\mathcal{E}|+|\mathcal{L}|)$. 

\begin{figure}[htb]
	\centering
	\includegraphics[width=0.48\textwidth]{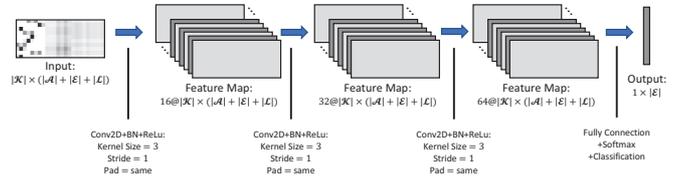}
	\caption{Designed CNN Architecture}
	\label{fig:CNN}
\end{figure}

We use the grayscale image as the feature to be learnt by CNN in step $iii$. The proposed CNN structure is illustrated in Figure \ref{fig:CNN}. It is worth noticing that the original problem is determining the position of each request cache content, i.e. $x_{ke}$. Similarity to \cite{lee2018deep}, for the aim of reducing the design complexity, the original caching assignment problem is decomposed into a number of independent sub-problems via the first order strategy \cite{zhang2013review}, which demonstrates that the CNN training can be processed in parallel. Each CNN corresponds to a related flow request, whose output is a vector representing the probability of each potential EC for a specific flow.
% Since each CNN is independent, training them in parallel benefits the computation complexity.

In step $iv$, we estimate the quality of CNN prediction with the optimal solution via cross entropy loss function:
\begin{equation}
\label{eq:loss}
    loss=-\sum_{i=1}^{|\mathcal{T}|}\sum_{e=1}^{|\mathcal{E}|}x^i_{ke}\ln{\hat{x}^i_{ke}},  k=1,2,\cdots, |\mathcal{K}|
\end{equation}
where $\mathcal{T}$ represents the training dataset, $x^i_{ke}$ is the training label of $i^{th}$ sample and $\hat{x}^i_{ke}$ is the prediction accordingly. CNN is trained recursively to reduce the value of loss function \eqref{eq:loss}.
% \textit{loss function of classification problem}
\subsection{\textcolor{black}{Testing Process}}
\begin{figure}[htb]
	\centering
	\includegraphics[width=0.48\textwidth]{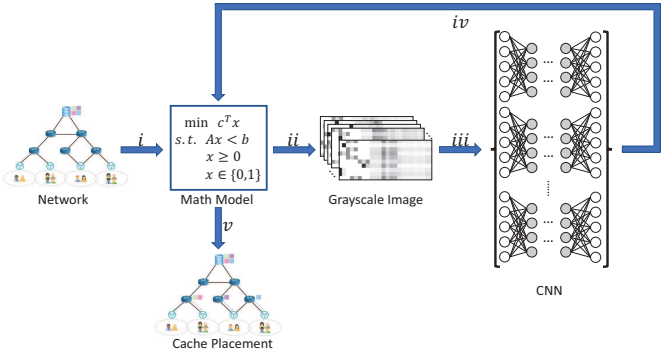}
	\caption{Testing Process}
	\label{fig:testing}
\end{figure}

Figure \ref{fig:testing} illustrates the testing process, where step $i$, $ii$ and $iii$ are same with training process in subsection \ref{sec:subtraining}.

In step $iv$, the output of each CNN is a vector where each element representing the confidence of prediction. For example, the output of a CNN is $[0.8,0,0.2,0,0,0]$, which means the CNN is more confident to put the content in EC1 whose value is 0.8 compared with caching in EC3 and the rest ECs are not in consideration. By combining these outputs together, we can get a matrix indicating the preference of caching locations. Based on the previous training progress, we believe it is a small-probability event that the optimal solution locates in these 0 value positions. For the purpose of reducing the number of non-zero elements in the matrix, a threshold $\delta$ is set for the predicted probability which force the value less than the threshold to be zero or to be one otherwise. Using a similar methodology with  \cite{lei2019learning}, the matrix $\mathcal{O}=\{o_{ke},\forall k\!\in\!\mathcal{K},e\!\in\!\mathcal{E}\}$ after threshold filter can be used to reduce the number of decision variables in optimization problem \eqref{LP:main} by adding the following new constraint 
\begin{equation}
\label{eq:added}
x_{ke}\leq o_{ke},\forall k\!\in\!\mathcal{K},e\!\in\!\mathcal{E}
\end{equation}

In step $v$ we can efficiently solve the MILP to obtain a caching assignment when a significant part of $\mathcal{O}$ is zero.

\section{Numerical Investigations}
\label{sec:investigations}
% Comparisons can be added:
% \begin{itemize}
%     \item problem compression after CNN, such as \# of variables and \# of constraints;
%     \item gain of computation time;
%     \item error/convergence during training with different parameter setting, such as layers or CNN architecture (refer to the criteria in Sensor paper);
%     \item could we use heatmap to express our results?
    % \item how to measure overfitting/underfitting.
% \end{itemize}

In this section, we evaluate the quality of performance among the solutions derived from the MILP, the proposed CNN and the Greedy Caching Algorithm (GCA) which attempts to assign each flow to its nearest available EC respectively. For the aim of comparing fairly, a penalty would be added to the total cost when constraints are invalided. %In that case, when a new assignment with smaller $TC^N$ is found, it  replace the original solution. We assume a mesh tree-like mobile network where user mobility is take placing on the edge of the network. Based on user mobility patterns  we apply the different techniques, i.e., MILP, CNN, GCA and RGC respectively, to efficiently perform pro-active edge cloud caching of popular content. The mobility of users between the different AR $a$ depends on the moving probability $P_{ka}$ and their requests for content. 
Table \ref{tab:Network_Parameters} provides a summary of the key parameters used for numerical investigations. 

\begin{table}[htb]
\centering
\caption{\label{tab:Network_Parameters} Key Network Parameters \cite{wang2019proactive}}
% \cite{wang2019caching}\cite{zheng2016optimal}}
\begin{tabular}{l|c}
\hline
\textbf{Parameter}&\textbf{value} \\
\hline
degree per node & 2$\sim$5 \\
mobile end users ($|\mathcal{K}|$)& \{5,10,15,20\} \\
number of links ($|\mathcal{L}|$)& 20\\
number of access routers ($|\mathcal{A}|$)& 7\\
number of edge clouds ($|\mathcal{E}|$)& 6\\
weight of hosting a cache  ($\alpha$)& [0,1] \\
weight of communication cost ($\beta$)& [0,1] \\
% factor of invalid constraint penalty ($\gamma$)& 20 \\
threshold of prediction probability ($\delta$) & 0.001 \\
end user request content size ($s_k$) & [10,50]MB \\
available cache size in EC ($w_e$) & [100,500]MB\\
user request transmission bandwidth ($b_k$) & [1,10]Mbps\\
link available capacity ($c_l$) & [50,100]Mbps\\
% number of epochs for RGC & 500 \\
\hline
\end{tabular}
\end{table}

The data set is constructed via solving the mathematical model \eqref{LP:main} by MILP and the samples are based on different scenarios which are represented by the combination of $p_{ka}$, $s_k$, $b_k$, $w_e$ and $c_l$. For the case of $|\mathcal{K}|=5$, we generate 1000 samples where $900$ are used for training and the rest for testing. In the training process, 5 CNNs are tamed independently with same input grayscale image but different outputs where each output corresponds to a specific mobile user request assignment. Then every $100$ samples are built for the scenario of 10, 15 and 20 users respectively to estimate the performance of different approaches, i.e. MILP, CNN and GCA. In order to use trained CNN solving these scenarios, we split the input grayscale image into several same size sub images whose height matches the number of batched CNNs which is 5 in this paper. After that trained CNNs are used for predicting the allocation of sub images and update the unassigned sub images. For instance, when reading a grayscale image representing 15 users, we separate them into 3 sub images whose size is 5 users. Then we call the trained batched CNNs to solve the first sub image and update the rest two depending on the preceding allocation by selecting the EC with the maximum predicted probability, such as updating the value of $q_{ke}$ via $$q_{ke}=\frac{s_k}{w_e-\sum_{k'\in\mathcal{K'}}s_{k'}x_{k'e}}, \forall k\in\mathcal{K}-\mathcal{K'}$$ where $\mathcal{K'}$ is the set of mobile users who have already been assigned. We keep doing that process recursively until all the users' requests are allocated.

In Table \ref{tab:Performance_Comparison}, the performances of these approaches are compared. The mean computation time can be viewed as the time complexity for these three algorithms. When it comes to CNN, time is tracked between loading the grayscale image and receiving the solution of optimization problem excluding the period used for training. Then we calculate the average total cost with invalided constraints penalty among these methods. Next, we collate the final precision of these solutions of 100 test samples which is calculated by $$precision=\frac{1}{100}\sum_{i=1}^{100}\frac{|X^i\cap\hat{X}^i|}{|\hat{X}^i|}$$ where $X^i$ is the optimal solution of $i^{th}$ sample and $\hat{X}^i$ is the solution of estimated algorithms. After that the feasible ratio are investigated which is the percentage of the constraints-satisfied assignments. In addition, the maximum total cost difference with the optimal solution MILP are analyzed, which represents the quality gap of solutions in the worst case. Finally, the average number of decision variables of MILP and CNN in the 100 testing samples are expressed\footnote{Simulations run on MATLAB 2019b in a 64-bit Windows 10 environment on a machine equipped with an Intel Core i7-7700 CPU 3.60 GHz Processor and 16 GB RAM}.

As shown in Table \ref{tab:Performance_Comparison}, the GCA can finish caching allocations within 0.1s in all scenarios, which is the fastest among the three methods. However, it also has the largest cost payment and the average cost gap with MILP increases from 39.6\% to 75.9\%. Additionally, the other criteria, such as feasible ratio and maximum difference with optimal solution, become worse with the increment of mobile users due to the penalty cost of infeasible constraints. From this aspect, GCA can be viewed as the lower-bound of caching assignment. On the other hand, MILP outperforms the rest two approaches with the increase of the number of requests and is the upper-bound which limits the ceiling of performance. The CNN can provide a competitive solution with less time complexity compared with MILP. In the scenario of 5 requests, the proposed CNN performs similarly as MILP where the cost gap is less than 3\textperthousand\ even in the worst case. Furthermore, the computation process is accelerated by 6 times because the number of decision variables is decreased by more than 34.4\%. Even in the case of 15 flow requests, CNN speed up the computation process by 75 times with less than 5\% loss on total cost. However, the calculation time of proposed algorithm becomes more than 1 hour in 20 requests since the number of decision variables 960.2 is still relatively large for an optimization problem even after CNN reduction. Moreover, the feasible ratio of CNN keeps 100\% in the testing. This is expected because the constraints in proposed algorithm is stricter than MILP with one more limitation \eqref{eq:added}.
\begin{table*}[htb]
\centering
\caption{\label{tab:Performance_Comparison}Performance Comparison.}
\begin{tabular}{l|c|c|c|c|c|c|c}
\hline
 & Methods & Computation Time & Mean TC & Precision & Feasible Ratio & Maximum Diff & number of decision variables\\
\hline
\hline
\multirow{3}{*}{5 requests} & \textbf{MILP} & 0.49s & 10.83 & 100.0\% & 100.0\% & - & 376.0\\
& \textbf{CNN} & 0.08s & 10.83 & 98.8\% & 100.0\% & 0.03 & 246.6 \\ 
& \textbf{GCA} & 0.01s & 17.95 & 63.5\% & 99.8\% & 67.20 & - \\
\hline
\multirow{3}{*}{10 requests} & \textbf{MILP} & 80.96s & 23.31 & 100.0\% & 100.0\% & - & 746.0 \\ & \textbf{CNN} & 0.85s & 24.32 & 95.4\% & 100.0\% & 1.90 & 484.7 \\
& \textbf{GCA} & 0.03s & 58.28 & 57.1\% & 95.2\% & 329.00 & - \\
\hline
\multirow{3}{*}{15 requests} & \textbf{MILP} & 1.58h & 39.04 & 100.0\% & 100.0\% & - & 1116.0 \\
& \textbf{CNN} & 74.94s & 41.05 & 76.8\% & 100.0\% & 13.70 & 727.9\\
& \textbf{GCA} & 0.04s & 124.75 & 51.9\% & 89.0\% & 463.90 & - \\
\hline
\multirow{3}{*}{20 requests} & \textbf{MILP} & 2.40h & 64.91 & 100.0\% & 100.0\% & - & 1486.0 \\
& \textbf{CNN} & 1.30h & 68.20 & 67.8\% & 100.0\% & 32.23 & 960.2 \\
& \textbf{GCA} & 0.08s & 269.46 & 44.4\% & 80.8\% & 646.10 & - \\
\hline
\end{tabular}
\end{table*}

\section{Conclusions}
\label{sec:conclusions}
AI-based data-driven techniques are attracting significant research attention in taming the complexity of 5G and beyond networks. This work presents a framework where a data-driven technique in the form of a deep convolutional neural network is amalgamated with a model based technique in the form of an integer programming optimization problem. The CNN is trained offline with optimal decisions for the purpose of reducing the number of decision variables so that to speed decision in order to be amenable for real time implementations. Numerical investigations reveal that in the case of 15 flow requests the proposed method managed to provide a speed up on the calculation of edge cache locations by 75 times compared to the MILP with a cost on the quality of the decision making of less than 5\%.

Leveraging on recent promising results in the area of artificial intelligence is a promising direction for taming the complexity in the operation of emerging and future 5G and beyond networks. To this end, there are various future avenues of research in the area of utilizing AI techniques to provide high quality decision making in real-time. Issues of reliability of the decision making and robustness of the solutions require significant attention.
%represent a powerful method to use for a wide variety of image processing tasks. Their  operational characteristics create a link between general deep feed-forward neural networks and adaptive filtering. 
%Inspired by these results in this paper we present an approach of transforming an optimization problem related to  caching of popular content on a number of edge clouds to a grayscale image so that to train deep convolutional neural networks.
%A wide range of numerical investigations provide insights about the capabilities of the proposed approach to successfully derive highly competitive policies for caching of popular content on different edge clouds. 
%Numerical investigations reveal that the proposed scheme can provide  real-time decision making which can be even more than 400\% better than powerful randomized greedy heuristics which are one of the options for real time decision making.  
% 
%Future avenues of research are multifaceted. Further analysis is needed in the area of sensitivity of the deep learning output to -  controlled  - small variations of the incoming requests. This can be viewed as sensitivity analysis as with respect to the integer linear programming formulation and adversarial behaviour of the CNN. Furthermore, an interesting extension would be to include temporal characteristics of the problem that will require new ways of transforming the spatio-temporal aspects of the optimization problem to an image.
\addtolength{\textheight}{-12cm}   
\bibliographystyle{ieeetr}
\bibliography{reference}

\begin{thebibliography}{10}

\bibitem{SurveyCaching}
S.~{Wang}, X.~{Zhang}, Y.~{Zhang}, L.~{Wang}, J.~{Yang}, and W.~{Wang}, ``A
  survey on mobile edge networks: Convergence of computing, caching and
  communications,'' {\em IEEE Access}, vol.~5, pp.~6757--6779, 2017.

\bibitem{wang2019proactive}
Y.~Wang, G.~Zheng, and V.~Friderikos, ``Proactive caching in mobile networks
  with delay guarantees,'' in {\em ICC 2019-2019 IEEE International Conference
  on Communications (ICC)}, pp.~1--6, IEEE, 2019.

\bibitem{zappone2019wireless}
A.~Zappone, M.~Di~Renzo, and M.~Debbah, ``Wireless networks design in the era
  of deep learning: Model-based, ai-based, or both?,'' {\em arXiv preprint
  arXiv:1902.02647}, 2019.

\bibitem{sun2019application}
Y.~Sun, M.~Peng, Y.~Zhou, Y.~Huang, and S.~Mao, ``Application of machine
  learning in wireless networks: Key techniques and open issues,'' {\em IEEE
  Communications Surveys \& Tutorials}, 2019.

\bibitem{sze2017efficient}
V.~Sze, Y.-H. Chen, T.-J. Yang, and J.~S. Emer, ``Efficient processing of deep
  neural networks: A tutorial and survey,'' {\em Proceedings of the IEEE},
  vol.~105, no.~12, pp.~2295--2329, 2017.

\bibitem{cheng2017mobile}
X.~Cheng, L.~Fang, L.~Yang, and S.~Cui, ``Mobile big data: The fuel for
  data-driven wireless,'' {\em IEEE Internet of things Journal}, vol.~4, no.~5,
  pp.~1489--1516, 2017.

\bibitem{chen2019artificial}
M.~Chen, U.~Challita, W.~Saad, C.~Yin, and M.~Debbah, ``Artificial neural
  networks-based machine learning for wireless networks: A tutorial,'' {\em
  IEEE Communications Surveys \& Tutorials}, 2019.

\bibitem{tanzil2017adaptive}
S.~S. Tanzil, W.~Hoiles, and V.~Krishnamurthy, ``Adaptive scheme for caching
  youtube content in a cellular network: Machine learning approach,'' {\em IEEE
  Access}, vol.~5, pp.~5870--5881, 2017.

\bibitem{lei2019learning}
L.~Lei, Y.~Yuan, T.~X. Vu, S.~Chatzinotas, and B.~Ottersten, ``Learning-based
  resource allocation: Efficient content delivery enabled by convolutional
  neural network,'' in {\em 2019 IEEE 20th International Workshop on Signal
  Processing Advances in Wireless Communications (SPAWC)}, pp.~1--5, IEEE,
  2019.

\bibitem{lee2018deep}
M.~Lee, Y.~Xiong, G.~Yu, and G.~Y. Li, ``Deep neural networks for linear sum
  assignment problems,'' {\em IEEE Wireless Communications Letters}, vol.~7,
  no.~6, pp.~962--965, 2018.

\bibitem{lei2017deep}
L.~Lei, L.~You, G.~Dai, T.~X. Vu, D.~Yuan, and S.~Chatzinotas, ``A deep
  learning approach for optimizing content delivering in cache-enabled
  hetnet,'' in {\em 2017 international symposium on wireless communication
  systems (ISWCS)}, pp.~449--453, IEEE, 2017.

\bibitem{nowak2018revised}
A.~Nowak, S.~Villar, A.~S. Bandeira, and J.~Bruna, ``Revised note on learning
  quadratic assignment with graph neural networks,'' in {\em 2018 IEEE Data
  Science Workshop (DSW)}, pp.~1--5, IEEE, 2018.

\bibitem{van2014performance}
P.~Van~Mieghem, {\em Performance analysis of communications networks and
  systems}.
\newblock Cambridge University Press, 2014.

\bibitem{vasilakos2012proactive}
X.~Vasilakos, V.~A. Siris, G.~C. Polyzos, and M.~Pomonis, ``Proactive selective
  neighbor caching for enhancing mobility support in information-centric
  networks,'' in {\em Proceedings of the second edition of the ICN workshop on
  Information-centric networking}, pp.~61--66, ACM, 2012.

\bibitem{zhang2013review}
M.-L. Zhang and Z.-H. Zhou, ``A review on multi-label learning algorithms,''
  {\em IEEE transactions on knowledge and data engineering}, vol.~26, no.~8,
  pp.~1819--1837, 2013.

\end{thebibliography}

\end{document}